\documentclass{article}
\usepackage{amsfonts}
\usepackage{amsmath}

\input{tcilatex}
\begin{document}

\title{\textbf{Gravitational Waves}\\
\textbf{and Loop Quantum Gravity}}
\author{W. F. Chagas-Filho \\
Physics Department\\
Federal University of Sergipe, SE, Brazil}
\maketitle

\begin{abstract}
Loop Quantum Gravity is a formalism for describing the quantum mechanics of
the gravitational field based on the canonical quantization of General
Relativity. The most important result of \ LQG is that geometric quantities
such as area and volume are not arbitrary but are quantized in terms of a
minimum length. In this paper we investigate the effect of a minimum length
in wave propagation. We find that the minimum length, combined with the
constancy of the speed of light, induces a quantization of the energy of a
gravitational wave.
\end{abstract}

\section{\protect\bigskip Introduction}

With the detection of gravitational waves in 2015 no more room was left for
doubts about the validity of General Relativity as the fundamental theory
for the gravitational interaction. The detection of gravitational waves also
confirms the point of view that has been advocated by some authors [1,2,3]
that the theory describing the quantum mechanics of the gravitational
interaction should be based on the quantization of General Relativity (GR).

General Relativity is based on the invariance of the theory under space-time
diffeomorphisms. As a consequence of this invariance, in GR the space and
time coordinates have no physical meaning. Physical predictions of GR are
independent of these coordinates. Only predictions based on relations
between observable quantities have a physical meaning in GR.

Today it is believed that the gravitational waves predicted by GR can
propagate in space-time only in weak gravitational fields. In this case GR
becomes a linear theory. In the linear theory we can expand the curved
space-time metric $g_{\mu \nu }(x)$ around the flat Minkowski metric $\eta
_{\mu \nu }$ according to 
\begin{equation}
g_{\mu \nu }(x)=\eta _{\mu \nu }+h_{\mu \nu }(x)  \tag{1}
\end{equation}%
where $h_{\mu \nu }(x)$ is a small deviation from the flat metric. A
necessary and crucial step to describe the propagation of gravitational
waves in the linear theory is to impose the Coulomb, or radiation gauge [4].
In Electrodynamics in flat space-time the Coulomb gauge is given by $%
\partial ^{a}A_{a}=0$, where $a=1,2,3,$ $\partial ^{a}$ is the ordinary
partial derivative associated to the Minkowski metric $\eta _{\mu \nu }$ and 
$A_{a}$ is the vector potential. In Electrodynamics in a curved space-time
with metric $g_{\mu \nu }(x)$ the Coulomb gauge is given by $D^{a}A_{a}=0$,
where $D^{a}$ is the covariant derivative associated to the metric $g_{\mu
\nu }(x)$. To reach the Coulomb gauge in linear GR, in terms of the small
deviation $h_{\mu \nu }(x)$ from the flat metric, is not an easy task. It
requires a detailed and careful control of the residual invariances of the
linear theory [4].

The most developed formalism for describing the quantum mechanics of the
gravitational field based on GR is called Loop Quantum Gravity (LQG). LQG is
based on the canonical quantization of GR using a connection formalism [5]
that can be constructed after the ADM [6] splitting of space-time into space
and time is performed. The most important result of LQG is that geometric
quantities, such as area and volume, can not have arbitrary values in a
gravitational field. Instead, areas and volumes are quantized in terms of a
minimum length [1,3] 
\begin{equation}
l_{0}=\sqrt{8\pi \gamma }L_{P}  \tag{2}
\end{equation}%
where $\gamma $ is the Immirzi parameter [7], a constant between 0 and 1
that fixes the precise scale of the quantum theory, and $L_{P}=1,62\times
10^{-35}m$ is the Planck length. The eigenvalues of the area operator, for
instance, are given in terms of this minimum length by [1,3] 
\begin{equation}
A_{j}=l_{0}^{2}\Sigma _{i}\sqrt{j_{i}(j_{i}+1)}  \tag{3}
\end{equation}%
where $j_{i}=\frac{1}{2},1,...$are the spins of the corresponding link. The
eigenvalues of the volume operator are also given in terms of the minimum
length $l_{0}$.

In a recent paper [8] it was shown that, by using a Hamiltonian duality
transformation in the phase space of LQG, it is possible to construct a
classical first-order formalism which yields a quantum theory that can be
interpreted as LQG in the momentum representation. One of the interesting
features of this classical formalism is that it has an $SU(2)$
generalization of the Coulomb gauge of Electrodynamics in curved space as
one of its basic equations [8]. In addition, in the quantum theory, the
generalized $SU(2)$ Coulomb gauge becomes a restriction on the wave
functionals of LQG in the momentum representation [8]. These results suggest
that gravitational waves can also propagate in gravitational fields which
are not weak. This fact has been confirmed in modern treatments of General
Relativity using numerical simulations, which indicate that gravitational
waves can also propagate in the nonlinear strong-gravity regime. For details
on this point see ref. [12], chapter 24.

Based on the results mentioned above, it becomes important to investigate
the possibility of combining the prediction of LQG about the existence of a
minimum length with the propagation of gravitational waves in nonlinear
strong gravitational fields. In this paper we present an initial step in the
direction of this investigation. For convenience we select a particular
value of the Immirzi parameter $\gamma $. The minimum eigenvalue of the area
operator occurs for $j=\frac{1}{2}$ and is given by 
\begin{equation}
A_{\min }=4\sqrt{3}\pi \gamma L_{P}^{2}  \tag{4}
\end{equation}%
We choose $\gamma =(4\sqrt{3}\pi )^{-1}$ which gives $A_{\min }=L_{P}^{2}$.
Therefore in this paper the minimum length that defines all physically
possible areas and volumes in a gravitational field is the Planck length $%
L_{P}$.

This paper is organized as follows. In section two we first briefly rewiew
the basic quantum equations of the usual formulation of LQG in the
configuration representation. The we briefly review the basic equations of a
quantum theory that can be interpreted as LQG in the momentum
representation. In section three we review the Planck scale. In section four
we obtain a quantization of the energy of a gravitational wave by using the
ideas of sections two and three. Concluding remarks appear in section five.

\section{Loop Quantum Gravity}

After the ADM splitting is performed, and the Ashtekar%
\'{}%
s connection variables $A_{a}^{i}(x)$ are introduced, GR can be cast as a
constrained Hamiltonian system with first-class [9] constraints. These
constraints are given by 
\begin{equation}
D_{a}E_{i}^{a}=0  \tag{5a}
\end{equation}%
\begin{equation}
F_{ab}^{i}E_{i}^{b}=0  \tag{5b}
\end{equation}%
\begin{equation}
F_{ab}^{ij}E_{i}^{a}E_{j}^{b}=0  \tag{5c}
\end{equation}%
Here $E_{i}^{a}(x)$ is the canonical momentum conjugated to the connection
variable $A_{a}^{i}(x)$, $i,j=1,2,3$ are internal $SU(2)$ indices and $%
a,b=1,2,3$ are space indices.%
\begin{equation}
D_{a}V^{i}=\partial _{a}V^{i}+\epsilon _{jk}^{i}A_{a}^{j}V^{k}  \tag{6}
\end{equation}%
defines the covariant derivative on the tangent space of a three dimensional
manifold $\Sigma $ without boundaries,%
\begin{equation}
F_{ab}^{i}=\partial _{a}A_{b}^{i}-\partial _{b}A_{a}^{i}+\epsilon
_{jk}^{i}A_{a}^{j}A_{b}^{k}  \tag{7}
\end{equation}%
and $F_{ab}^{ij}=\epsilon _{k}^{ij}F_{ab}^{k}$. Constraint (5a) is the Gauss
law constraint. It generates internal $SU(2)$ gauge transformations.
Constraint (5b) generates space diffeomorphisms and constraint (5c) is the
Hamiltonian constraint. It can be shown [1] that constraints (5) are
equivalent to Einstein%
\'{}%
s equation.

The transition to Loop Quantum Gravity is performed by promoting the
canonical momenta $E_{i}^{a}$ to quantum operators $\hat{E}_{i}^{a}=-i$%
h{\hskip-.2em}\llap{\protect\rule[1.1ex]{.325em}{.1ex}}{\hskip.2em}%
$\frac{\delta }{\delta A_{a}^{i}}$ which act on configuration space wave
functionals $\Psi (A)$. The constraints (5) then become the quantum
equations 
\begin{equation}
D_{a}\frac{\delta }{\delta A_{a}^{i}}\Psi (A)=0  \tag{8a}
\end{equation}%
\begin{equation}
F_{ab}^{i}\frac{\delta }{\delta A_{b}^{i}}\Psi (A)=0  \tag{8b}
\end{equation}%
\begin{equation}
F_{ab}^{ij}\frac{\delta }{\delta A_{a}^{i}}\frac{\delta }{\delta A_{b}^{j}}%
\Psi (A)=0  \tag{8c}
\end{equation}%
Equations (8) are the basic equations of LQG in the configuration
representation [1]. When these equations are valid the space has a minimum
length $L_{P}$ (for our choice of the Immirzi parameter $\gamma $).

On the other hand, the formal dual action for GR introduced in [8] gives the
constraints%
\begin{equation}
D^{a}A_{a}^{i}=0  \tag{9a}
\end{equation}%
\begin{equation}
F_{i}^{ab}A_{b}^{i}=0  \tag{9b}
\end{equation}%
\begin{equation}
F_{ij}^{ab}A_{a}^{i}A_{b}^{j}=0  \tag{9c}
\end{equation}%
where now 
\begin{equation}
D^{a}V^{i}=\partial ^{a}V^{i}+\epsilon ^{ijk}E_{j}^{a}V_{k}  \tag{10}
\end{equation}%
defines the covariant derivative on the cotangent space of $\Sigma $,%
\begin{equation}
F_{i}^{ab}=\partial ^{a}E_{i}^{b}-\partial ^{b}E_{i}^{a}+\epsilon
_{i}^{jk}E_{j}^{a}E_{k}^{b}  \tag{11}
\end{equation}%
and $F_{ij}^{ab}=\epsilon _{ijk}F_{k}^{ab}$. Notice that constraint (9a) is
a gravitational $SU(2)$ generalization of the Coulomb gauge of
Electrodynamics in curved spaces. While in the weak gravitational field
limit of usual GR a considerable amount of work is required to reach the
Coulomb gauge (a necessary step to describe gravitational waves) here we
have a dual formulation of GR in terms of connection variables in which the
Coulomb gauge is one of its basic equations. Since the weak gravitational
field limit was not required to arrive at the generalized $SU(2)$ Coulomb
gauge (9a) we interpret this as an indication that gravitational waves can
also propagate in gravitational fields which are not weak. As we mentioned
in the introduction, the fact that this interpretation is correct has
already been confirmed in modern treatments of GR [12].

Now the transition to the quantum theory is performed by promoting the
connection variables $A_{a}^{i}$ to quantum operators $\hat{A}_{a}^{i}=i$%
h{\hskip-.2em}\llap{\protect\rule[1.1ex]{.325em}{.1ex}}{\hskip.2em}%
$\frac{\delta }{\delta E_{i}^{a}}$ which act on momentum space wave
functionals $\Psi (E).$ The constraints (9) then become the quantum equations%
\begin{equation}
D^{a}\frac{\delta }{\delta E_{i}^{a}}\Psi (E)=0  \tag{12a}
\end{equation}%
\begin{equation}
F_{i}^{ab}\frac{\delta }{\delta E_{i}^{b}}\Psi (E)=0  \tag{12b}
\end{equation}%
\begin{equation}
F_{ij}^{ab}\frac{\delta }{\delta E_{i}^{a}}\frac{\delta }{\delta E_{j}^{b}}%
\Psi (E)=0  \tag{12c}
\end{equation}%
The quantum equations (12) are interpreted as defining Loop Quantum Gravity
in the momentum representation [8].

The quantum equations (8) and (12) are complementary and can be assumed to
be simultaneously valid. When this is done, it opens the new possibility
that gravitational waves can propagate in gravitational fields which are not
weak and which are defined on spaces with a minimum length.

\section{The Planck scale}

In this section we review the Planck scale. We start by considering the
three fundamental constants of physics:

1) the speed of light in empty space 
\begin{equation}
c=2,998\times 10^{8}m/s  \tag{13}
\end{equation}

2) Newton%
\'{}%
s gravitational constant 
\begin{equation}
G=6,67\times 10^{-11}N.m^{2}.Kg^{-2}  \tag{14}
\end{equation}

3) Planck%
\'{}%
s constant 
\begin{equation}
\text{%
h{\hskip-.2em}\llap{\protect\rule[1.1ex]{.325em}{.1ex}}{\hskip.2em}%
=}\frac{h}{2\pi }=1,055\times 10^{-34}J.s  \tag{15}
\end{equation}

In 1899 Planck [10] noticed that by combining these fundamental constants in
a unique way he could define a fundamental scale of time, length and mass.
Today this fundamental scale is called the Planck scale [11]. It is given by 

1) the Planck time%
\begin{equation}
T_{P}=\sqrt{\frac{\text{%
h{\hskip-.2em}\llap{\protect\rule[1.1ex]{.325em}{.1ex}}{\hskip.2em}%
G}}{c^{5}}}=5,40\times 10^{-44}s  \tag{16}
\end{equation}

2) the Planck length 
\begin{equation}
L_{P}=\sqrt{\frac{\text{%
h{\hskip-.2em}\llap{\protect\rule[1.1ex]{.325em}{.1ex}}{\hskip.2em}%
G}}{c^{3}}}=1,62\times 10^{-35}m  \tag{17}
\end{equation}

3) the Planck mass 
\begin{equation}
M_{P}=\sqrt{\frac{\text{%
h{\hskip-.2em}\llap{\protect\rule[1.1ex]{.325em}{.1ex}}{\hskip.2em}%
c}}{G}}=2,17\times 10^{-5}g  \tag{18}
\end{equation}%
Notice that the Planck length is the distance that light travels during the
Planck time. Therefore the existence of a minimum length $L_{P}$, together
with the constancy of the speed of light $c$, automatically leads to the
existence of a minimum time $T_{P}$.

\section{Gravitational waves and Loop Quantum Gravity}

As we saw in section two, combining the basic equations of LQG in the
configuration and in the momentum representation opens the new possibility
that gravitational waves can propagate in gravitational fields which are not
weak and which are defined on spaces with a minimum length equal to the
Planck length. We may then expect that the wave length of such gravitational
waves should be an integer multiple of $L_{P}$, that is

\begin{equation}
\lambda _{n}=nL_{P}\text{ \ \ \ \ \ \ \ \ \ \ }n=1,2,...  \tag{19}
\end{equation}%
As we saw in section three, the existence of a minimum length $L_{P}$
combined with the constancy of the speed of light leads to the existence of
a minimum time $T_{P},$ which is the time necessary for ligth to travel the
distance $L_{P}$. We may then expect that the period of such gravitational
waves should be an integer multiple of the Planck time%
\begin{equation}
P_{n}=nT_{P}\text{ \ \ \ \ \ \ \ \ }n=1,2,...  \tag{20}
\end{equation}%
Equations (19) and (20) have no physical meaning because, as we mentioned in
the introduction, space and time have no intrinsic physical meaning in GR.
For our purposes we need to find an observable physical quantity associated
with the gravitational wave. One such physical quantity is the energy of the
gravitational wave. Since electromagnetic waves and gravitational waves
propagated with the same speed of light $c$, we may assume that Planck%
\'{}%
s equation $E=h\nu $ for the energy of an electromagnetic wave with
frequency $\nu $ can also be applied in the case of gravitational waves.

Combining Planck%
\'{}%
s equation%
\begin{equation}
E=h\nu   \tag{21}
\end{equation}%
with equation (20), we can then write for a gravitational wave 
\begin{equation}
E_{n}=h\nu _{n}=h\frac{1}{P_{n}}=h\frac{1}{nT_{P}}\text{ \ \ \ \ \ \ }%
n=1,2,...  \tag{22}
\end{equation}%
Therefore we obtain%
\begin{equation}
E_{n}=\frac{1}{n}\sqrt{\frac{2\pi h^{2}c^{5}}{hG}}  \tag{23}
\end{equation}%
which can be rewritten as 
\begin{equation}
E_{n}=\frac{2\pi c^{2}}{n}\sqrt{\frac{\text{%
h{\hskip-.2em}\llap{\protect\rule[1.1ex]{.325em}{.1ex}}{\hskip.2em}%
c}}{G}}=\frac{2\pi }{n}E_{P}\text{ \ \ \ \ \ \ \ \ }n=1,2,...  \tag{24}
\end{equation}%
where we have introduced the Planck energy $E_{P}=M_{P}c^{2}$ with $M_{P}$
the Planck mass given in equation (18). We have therefore obtained a
quantization of the energy of a gravitational wave by combining the
existence of the minimum length $L_{P}$ with the constancy of the speed of
light $c$.

\section{Concluding remarks}

In this paper, based on what can be considered as two complementary
formulations of Loop Quantum Gravity, we gave a theoretical evidence that
gravitational waves can propagate in gravitational fields which are not
weak. This evidence is a gravitational $SU(2)$ generalization of the
Coulomb, or radiation gauge, of Electrodynamics in curved space-time. Using
the notion of a minimum length predicted by LQG and assuming the validity of
Planck%
\'{}%
s equation $E=h\nu $ in the case of gravitational waves, we obtained a
quantization of the energy of a gravitational wave in terms of the Planck
energy $E_{P}=M_{P}c^{2}$, where $M_{P}$ is the Planck mass.

\end{document}